%% file: Longer_version_BCpaper_v18.tex
\documentclass[conference]{IEEEtran}
\usepackage{amsmath,cite,amsfonts,amssymb,psfrag,amsthm}
\usepackage{graphicx}
\usepackage{epstopdf}
\usepackage{rotating}
\usepackage{dsfont}
\usepackage{color}
\usepackage{tikz}
\usetikzlibrary{plotmarks}
\usepackage{pgfplots}
\usetikzlibrary{calc}%,fillbetween
\usetikzlibrary{shapes,arrows}
\usetikzlibrary{decorations.markings}
\usetikzlibrary{positioning}
\pgfplotsset{compat=1.10}
\usetikzlibrary{calc}%,fillbetween
\usetikzlibrary{shapes,arrows}
\usetikzlibrary{decorations.markings}

\usepackage[utf8]{inputenc}
\usepackage{authblk}

\DeclareFontFamily{U}{mathx}{\hyphenchar\font45}
\DeclareFontShape{U}{mathx}{m}{n}{<-> mathx10}{}
\DeclareSymbolFont{mathx}{U}{mathx}{m}{n}
\DeclareMathAccent{\widebar}{0}{mathx}{"73}

\makeatletter
\newtheorem*{rep@theorem}{\rep@title}
\newcommand{\newreptheorem}[2]{%
	\newenvironment{rep#1}[1]{%
		\def\rep@title{\Cref{##1}}%
		\begin{rep@theorem}}%
		{\end{rep@theorem}}}
\newcommand*{\textlabel}[2]{%
	\edef\@currentlabel{#1}% Set target label
	\phantomsection% Correct hyper reference link
	#1\label{#2}% Print and store label
}
\makeatother

\newtheorem{theorem}{Theorem}

\newtheorem{remark}{Remark}
\newtheorem{definition}{Definition}

\newtheorem{lemma}{Lemma}

\begin{document}
\title{Private Authentication with Physical Identifiers Through Broadcast Channel Measurements}
\IEEEoverridecommandlockouts

  \author[1]{Onur G\"unl\"u}
 \author[1]{Rafael F. Schaefer}
 \author[2]{Gerhard Kramer}

 \affil[1]{Information Theory and Applications Chair, Technische Universit\"{a}t Berlin\\ \authorcr
 	\{guenlue, rafael.schaefer\}@tu-berlin.de}
 \affil[2]{Chair of Communications Engineering, Technical University of Munich\\ \authorcr
 	gerhard.kramer@tum.de}

\maketitle

%%%%%%%%%%%%%%%%%%%%%%%%%%%%%%%%%%%%%%%%%%%%%%%%%%%%%%%%%%%%%%%%%%%%%%%%%%%%%%%%%%%%%%%%%%%%%%%%%%%%%%%%%%%%%%%%%%%
\begin{abstract}
	A basic model for key agreement with biometric or physical identifiers is extended to include measurements of a hidden source through a general broadcast channel (BC). An inner bound for strong secrecy, maximum key rate, and minimum privacy-leakage and database-storage rates is proposed. The inner bound is shown to be tight for physically-degraded and less-noisy BCs.	 
\end{abstract}
%%%%%%%%%%%%%%%%%%%%%%%%%%%%%%%%%%%%%%%%%%%%%%%%%%%%%%%%%%%%%%%%%%%%%%%%%%%%%%%%%%%%%%%%%%%%%%%%%%%%%%%%%%%%%%%%%%%

\IEEEpeerreviewmaketitle
% %%%%%%%%%%%%%%%%%%%%%%%%%%%%%%%%%%%%%%%%%%%%%%%%%%%%%%%%%%%%%%%%%%%%
% %%%%
% %%%%               Introduction
% %%%%
% %%%%%%%%%%%%%%%%%%%%%%%%%%%%%%%%%%%%%%%%%%%%%%%%%%%%%%%%%%%%%%%%%%%%
\section{Introduction} \label{sec:intro}
Secret key generation from biometric or physical identifiers such as fingerprints, uncontrollable fine variations of ring oscillator (RO) outputs, or random start-up values of static random access memories (SRAMs) is a promising alternative to key storage in a non-volatile memory (NVM) \cite{GassendThesis}. Physical identifiers for digital devices, such as Internet-of-Things (IoT) devices, can be implemented using physical unclonable functions (PUFs) \cite{GassendThesis}. One can also use PUFs at the transmitter as a source of local randomness \cite[Chapter 1]{benimdissertation}. For instance, consider the wiretap channel \cite{WTC} where a transmitter sends a message through a broadcast channel (BC) \cite{CoverandThomas} so that a legitimate receiver can reliably decode the message, while the message is hidden from an eavesdropper by using a randomized encoder with random bits supplied by a PUF.

We use the source model for key agreement from \cite{AhlswedeCsiz,Maurer} to characterize rate regions for PUFs. Based on a source observation, an encoder generates a key and sends \textit{helper data} to a decoder, so the key agreement at the decoder is successful. The helper data can be made public as long as the information leaked about the secret key, i.e., \emph{secrecy leakage}, is negligible and the information leaked about the identifier output, i.e., \emph{privacy leakage}, is small \cite{IgnaTrans,LaiTrans}. The amount of storage should also be kept small to limit the hardware cost \cite{csiszarnarayan}.

Suppose the encoder generates a key from a noisy identifier output and a decoder has access to another noisy measurement of the same identifier and the helper data to reconstruct the key. We call this model the \emph{generated-secret} (GS) model. Similarly, for the \emph{chosen-secret} (CS) model, an embedded key and noisy identifier measurements are combined to generate the public helper data. We consider both models to address different applications.

The source, noisy identifier, and measurement symbol strings are related by a BC with one input and two outputs. In \cite{bizimMMMMTIFS,bizimKittipongTIFS}, the BC consists of two separate measurement channels. We extend this model to capture the effects of correlated noise in the measurements. As an example for this case, RO oscillation frequencies depend on the surrounding hardware logic, so the encoder and decoder measurements of the same RO PUFs tend to be correlated \cite{MerliROCorrelated}. 

We derive achievable key-leakage-storage rate tuples for a general BC with strong secrecy. The separate-measurement case in \cite{bizimKittipongTIFS,bizimMMMMTIFS} corresponds to a physically-degraded (PD) BC, and the visible source model in \cite{IgnaTrans,LaiTrans} corresponds to a semi-deterministic BC. The rate region for another PD BC scenario is given in this work, which does not follow from previous results due to the asymmetry in the constraints. We further establish the rate regions for less-noisy (LN) BCs and derive results with strong privacy and strong secrecy. We remark that in \cite{IgnaTrans,LaiTrans,MatthieuPolar}, a ``private" key that is available to the encoder and decoder and that is hidden from the eavesdropper is considered to obtain strong privacy. This assumption is not realistic since a private key requires hardware protection against physical attacks, and if such a protection is feasible then it is not necessary to use a PUF for key generation. 

This paper is organized as follows. In Section~\ref{sec:problem_setting}, we describe our models and the problem. We give achievable key-leakage-storage regions for the GS and CS models in Section~\ref{sec:achievablescheme}. The proposed inner bounds are shown in Section~\ref{sec:tightregions} to be tight for classes of PD and LN BCs. We give an example in Section~\ref{sec:example} to illustrate the effects of delay and hardware cost on the achieved rate tuples for a practical PUF design. Achievability proofs for the inner bounds and converses for a LN case are given in Sections~\ref{sec:achinner} and \ref{sec:convessentiallylessnoisycases}, respectively.

\section{Problem Definitions}\label{sec:problem_setting}
%%%%%%%%%%%%%%%%%%%%%%%%%%%%%%%%%%%%%%%%%%%%%%%%%%%%%%%%%%%%%%%%%%%%
%%%%%%%%%%%%%%%%%%%      SUBSECTION: PROBLEM I
%%%%%%%%%%%%%%%%%%%%%%%%%%%%%%%%%%%%%%%%%%%%%%%%%%%%%%%%%%%%%%%%%%%
Consider hidden identifier outputs $X^n$ that are independent and identically distributed (i.i.d.) with respect to a probability distribution $P_X$. The encoder and decoder observe noisy source measurements that are outputs of a BC $P_{\widetilde{X}Y|X}$. Suppose all alphabets are finite sets.

\begin{figure}
	\centering
	\resizebox{0.85\linewidth}{!}{
		\begin{tikzpicture}
		% Source Box
		\node (so) at (-1.5,-2.5) [draw,rounded corners = 5pt, minimum width=0.4cm,minimum height=0.6cm, align=left] {$P_X$};
		% Enc Box
		\node (a) at (1.5,0) [draw,rounded corners = 6pt, minimum width=2.10cm,minimum height=1.0cm, align=left] {$
		(S,W) \overset{(a)}{=} f_{1}^{(n)}(\widetilde{X}^n)$\\ $W\! \overset{(b)}{=}\! f_{2}^{(n)}(\widetilde{X}^n,S)$};
		%Key Database Box
		\node (kdb) at (1.5,3.6) [draw,rounded corners = 6pt, minimum width=1.0cm,minimum height=1.0cm, align=left] {$\;\;$ Key\\ Database};
		%Helper Data Database Box
		\node (hddb) at (3.9,2.9) [draw,rounded corners = 6pt, minimum width=1.0cm,minimum height=1.0cm, align=left] {$\;\;$ Public\\ Database};	
		%Comparison Box
		\node (comp) at (6.2,2.65) [draw,rounded corners = 6pt, minimum width=0.6cm,minimum height=0.6cm, align=left] {$=$};
		% S exchanges with Enc
		\node (quest) [right of = comp, node distance = 1.1cm] {?};		
		% Channel Box Xtilde
		\node (f) at (1.5,-2.5) [draw,rounded corners = 5pt, minimum width=1.2cm,minimum height=0.9cm, align=left] {$P_{\widetilde{X}Y|X}$};
		% Dec Box
		\node (b) at (6,0) [draw,rounded corners = 6pt, minimum width=3.2cm,minimum height=1.1cm, align=left] {$\hat{S} = g^{(n)}\left(Y^n,W\right)$};
		% X^N goes to Enc Box
		\node (a1) [right of = so, node distance = 1.35cm] {$X^n$};
		\node (b1) [below of = b, node distance = 2.5cm] {$Y^n$};
		% S exchanges with Enc
		\node (a2) [above of = a, node distance = 1.6cm] {$S$};
		% S exchanges with Enc
		\node (svhat) [right of = kdb, node distance = 4.3cm] {$S$};		
		% W_m
		\node (w5) [right of = a2, node distance = 2.40cm] {$W$};
		% S exchanges with Enc
		\node (shat) [right of = w5, node distance = 2.0cm] {$\hat{S}$};				
		% Source to X^N arrow
		\draw[decoration={markings,mark=at position 1 with {\arrow[scale=1.5]{latex}}},
		postaction={decorate}, thick, shorten >=1.4pt] (so.east) -- (a1.west);
		% X^N  to Channel Xtilde arrow
		\draw[decoration={markings,mark=at position 1 with {\arrow[scale=1.5]{latex}}},
		postaction={decorate}, thick, shorten >=1.4pt] (a1.east) -- (f.west);
		% \widetilde{X}^N  to Encoder arrow
		\draw[decoration={markings,mark=at position 1 with {\arrow[scale=1.5]{latex}}},
		postaction={decorate}, thick, shorten >=1.4pt] (f.north) -- (a.south) node [midway, right] {$\widetilde{X}^n$};
		% Channel box to Y^N arrow
		\draw[decoration={markings,mark=at position 1 with {\arrow[scale=1.5]{latex}}},
		postaction={decorate}, thick, shorten >=1.4pt] (f.east) -- (b1.west);
		% Y^N  to Enc arrow
		\draw[decoration={markings,mark=at position 1 with {\arrow[scale=1.5]{latex}}},
		postaction={decorate}, thick, shorten >=1.4pt] (b1.north) -- (b.south);
		% Enc to W 
		\draw[decoration={markings,mark=at position 1 with {\arrow[scale=1.5]{latex}}}, postaction={decorate}, thick, shorten >=1.4pt] (a.east) -- ($(b.west)-(0.87,0.01)$) -|($(w5.south)-(0.2,0)$);
		% Helper Data Database to Dec
		\draw[decoration={markings,mark=at position 1 with {\arrow[scale=1.5]{latex}}}, postaction={decorate}, thick, shorten >=1.4pt] ($(w5.south)+(0.18,0)$) -- ($(b.west)-(0.32,0.01)$)|- (b.west);
		% W to Helper Data Database 
		\draw[decoration={markings,mark=at position 1 with {\arrow[scale=1.5]{latex}}},
		postaction={decorate}, thick, shorten >=1.4pt] ($(w5.north)-(0.2,0)$) -- ($(hddb.south)-(0.2,0)$);
	    % Helper Data Database to W
	    \draw[decoration={markings,mark=at position 1 with {\arrow[scale=1.5]{latex}}},
	    postaction={decorate}, thick, shorten >=1.4pt] ($(hddb.south)+(0.2,0)$) -- ($(w5.north)+(0.2,0)$);
		% S to Enc arrow
		\draw[decoration={markings,mark=at position 1 with {\arrow[scale=1.5]{latex}}},
		postaction={decorate}, thick, shorten >=1.4pt]  ($(a2.south)+(0.2,0)$)-- ($(a.north)+(0.2,0)$) node [midway, right] {$(b)$};
		% Enc to S arrows
		\draw[decoration={markings,mark=at position 1 with {\arrow[scale=1.5]{latex}}},
		postaction={decorate}, thick, shorten >=1.4pt] ($(a.north)-(0.2,0)$)-- ($(a2.south)-(0.2,0)$) node [midway, left] {$(a)$};
		% S to  arrows
		\draw[decoration={markings,mark=at position 1 with {\arrow[scale=1.5]{latex}}},
		postaction={decorate}, thick, shorten >=1.4pt] ($(a2.north)-(0.2,0)$)-- ($(kdb.south)-(0.2,0)$) node [midway, left] {$(a)$};
		% Key Database to S arrow
		\draw[decoration={markings,mark=at position 1 with {\arrow[scale=1.5]{latex}}},
		postaction={decorate}, thick, shorten >=1.4pt]  ($(kdb.south)+(0.2,0)$)-- ($(a2.north)+(0.2,0)$) node [midway, right] {$(b)$};
		% Key Database to S_m arrow
		\draw[decoration={markings,mark=at position 1 with {\arrow[scale=1.5]{latex}}} ,
		postaction={decorate}, thick, shorten >=1.4pt]  ($(kdb.east)+(0,0)$)-- ($(svhat.west)+(0,0)$);
				% Dec to Shat arrow
				\draw[decoration={markings,mark=at position 1 with {\arrow[scale=1.5]{latex}}},
				postaction={decorate}, thick, shorten >=1.4pt]  ($(b.north)-(0.1,0)$)-- ($(shat.south)+(0,0)$);
				% Shat to Comparison arrow
				\draw[decoration={markings,mark=at position 1 with {\arrow[scale=1.5]{latex}}},
				postaction={decorate}, thick, shorten >=1.4pt]  ($(shat.north)+(0,0)$)-- (comp.south);
				% Svhat to Comparison arrow
				\draw[decoration={markings,mark=at position 1 with {\arrow[scale=1.5]{latex}}},
				postaction={decorate}, thick, shorten >=1.4pt]  ($(svhat.south)-(0,0)$)-- (comp.north);
				% Comparison to ? arrow
				\draw[decoration={markings,mark=at position 1 with {\arrow[scale=1.5]{latex}}},
				postaction={decorate}, thick, shorten >=1.4pt]  (comp.east) -- (quest.west);
		\end{tikzpicture}
	}
	\caption{Encoder and decoder measurements through a BC with $(a)$ the GS model and $(b)$ the CS model.}\label{fig:hiddengenerated}
\end{figure}
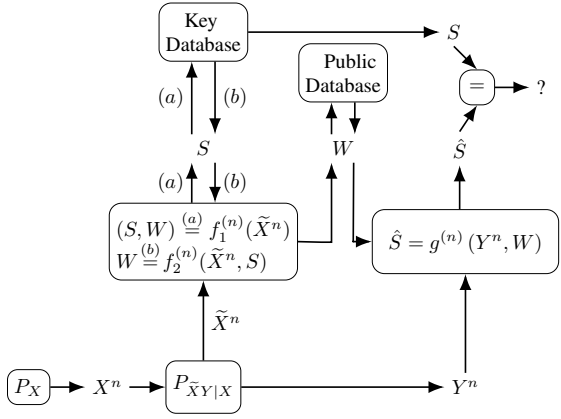

For the GS model shown in Fig.~\ref{fig:hiddengenerated}$(a)$, the encoder generates the helper data $W$ and a secret key $S$ from $\widetilde{X}^n$. The keys are stored in a secure database and the helper data are stored in a public database so that an eavesdropper has access only to the helper data. This is the case for PUFs when the helper data are stored without hardware protection. The eavesdropper does not observe a random sequence correlated with the identifier output. This is a valid assumption because for many physical and biometric identifiers, invasive attacks would permanently change the identifier outputs \cite{PappuThesis}. Furthermore, there are algorithms such as in \cite{bizimMDPI} to obtain almost i.i.d. outputs measured through memoryless channels. 

Using $W$ and the BC measurements $Y^n$, the decoder generates the key estimate $\hat{S}$ that is used for device authentication. Similar steps are applied for the CS model in Fig.~\ref{fig:hiddengenerated}$(b)$, except that $S$ is embedded into the encoder.

%%%%%%%%%%%%%%%%%%%%%%%%%%%%%%%%%%%%%%%%%%%%%%%%%%
%%%%%%%%%%%%%%%%%%%%%%%%%%%%%%%%%%%%%%%%%%%%%%%%%%%%%%%%%%%%%%%%%

\begin{definition}
	\normalfont A $\big(|\mathcal{W}^{(n)}|, |\mathcal{S}^{(n)}|,n\big)$-code $\mathcal{C}_n$ for key agreement with a hidden source with BC encoder and decoder measurements consists of
	\begin{itemize}
		\item[$\bullet$] a GS model encoder $f_{1}^{(n)}: \widetilde{\mathcal{X}}^{n}\rightarrow \mathcal{W}^{(n)}\times\mathcal{S}^{(n)}$ or\\ a CS model encoder $f_{2}^{(n)}: \widetilde{\mathcal{X}}^{n}\times\mathcal{S}^{(n)}\rightarrow \mathcal{W}^{(n)}$,
		\item[$\bullet$] a decoder $g^{(n)}: \mathcal{W}^{(n)}\times\mathcal{Y}^{n}\rightarrow \mathcal{S}^{(n)}$. \hfill$\lozenge$
	\end{itemize}
\end{definition}

\begin{definition}
\normalfont A (secret-key, privacy-leakage, storage), or (key-leakage-storage), rate tuple $(R_{s}, R_\ell,R_{w})$ is achievable for the GS and CS models with encoder and decoder measurements through the BC $P_{\widetilde{X}Y|X}$ if, given any $\delta\!>\!0$, there is some $n\!\geq\!1$, and an encoder and a decoder for which $\displaystyle R_{s}=\frac{\log|\mathcal{S}|}{n}$ and
\begin{alignat}{2}
		&\Pr[S\ne\hat{S}] \leq \delta &&\qquad\quad (reliability)\label{eq:reliabilityconst}\\
		&I(S;W) \leq \delta &&\qquad\quad(strong\; secrecy) \label{eq:secrecyconst}\\
		&\frac{1}{n}I(X^n;W)\! \leq\! R_{\ell}\!+\!\delta  &&\qquad\quad (privacy) \label{eq:privacyconst}\\
		&\frac{1}{n}H(S) \geq R_{s}-\delta  \quad&&\qquad\quad (uniformity)\label{eq:uniformityconst}\\
		&\frac{1}{n} \log|\mathcal{W}| \le R_{w}+\delta &&\qquad\quad (storage)\label{eq:storageconst}.
\end{alignat}
The \emph{key-leakage-storage} regions $\mathcal{C}_{\text{gs}}$ for the GS model and $\mathcal{C}_{\text{cs}}$ for the CS model are the closures of the set of all achievable rate tuples.\hfill $\lozenge$
\end{definition}

%%%%%%%%%%%%%%%%%%%%%%%%%%%%%%%%%%%%%%%%%%%%%%%%%%%%%%%%%%%%%%%%%%%%%%%%%%%%%%%%%%%%%%%%%%%%%%%%
%%%%%%%%%%%%%%%%%%%%%%%%%%%%%%%%%%%%%%%%%%%%%%%%%%%%%%%%%%%%%%%%%%%%%%%%%%%%%%%%%%%%%%%%%%%%%%%%
%%%%%%%%%%%%%%%%%%%%            SECTION III Key-storage-leakage-cost Regions  %%%%%%%%%%%%%%%%%%
%%%%%%%%%%%%%%%%%%%%%%%%%%%%%%%%%%%%%%%%%%%%%%%%%%%%%%%%%%%%%%%%%%%%%%%%%%%%%%%%%%%%%%%%%%%%%%%%
%%%%%%%%%%%%%%%%%%%%%%%%%%%%%%%%%%%%%%%%%%%%%%%%%%%%%%%%%%%%%%%%%%%%%%%%%%%%%%%%%%%%%%%%%%%%%%%%
\section{An Inner Bound}\label{sec:achievablescheme}
We are interested in characterizing the optimal trade-off among the secret-key, privacy-leakage, and storage rates with strong secrecy for BC measurements at the encoder and decoder. We give achievable rate regions for the GS and CS models in Theorem~\ref{theo:correlatedgscs}. The proofs are given Section~\ref{sec:achinner}.
\begin{theorem}[Inner Bounds for the GS and CS Models]\label{theo:correlatedgscs}
	An achievable rate region for the GS model is
	\begin{align}
	\mathcal{R}_{\text{gs}}\! =\! \bigcup_{P_{U|\widetilde{X}}}\!\Big\{&\left(R_s,R_\ell,R_w\right)\!\colon\!\nonumber\\
	&0\leq R_{s} \leq I(U;Y),\label{eq:corrkeyrate}\\
	&R_\ell \geq \max\{0,\; I(U;X)-I(U;Y)\}, \label{eq:corrleakagerate}\\
	&R_{w}\! \geq\! I(U;\widetilde{X})-I(U;Y) \label{eq:corrstoragerate}\Big\}
	\end{align}
	and an achievable rate region for the CS model is
	\begin{align}
	\mathcal{R}_{\text{cs}}\! =\! \bigcup_{P_{U|\widetilde{X}}}\!\Big\{&\left(R_s,R_\ell,R_w\right)\!\colon\!\nonumber\\
	&0\leq R_{s} \leq I(U;Y),\label{eq:cscorrkeyrate}\\
	&R_\ell \geq \max\{0,\; I(U;X)-I(U;Y)\},  \label{eq:cscorrleakagerate}\\
	&R_{w}\! \geq\! I(U;\widetilde{X}) \label{eq:cscorrstoragerate}\Big\}
	\end{align}
	where $U-\widetilde{X}-XY$ forms a Markov chain for $\mathcal{R}_{\text{gs}}$ and $\mathcal{R}_{\text{cs}}$.
\end{theorem}

Let $\mathcal{R}_{1}$ and as $\mathcal{R}_{2}$ be the rate regions characterized by (\ref{eq:corrkeyrate})-(\ref{eq:corrstoragerate}) when $\max\{0,\; I(U;X)-I(U;Y)\}$ is positive and zero, respectively. One can limit the cardinality of $U$ to $|\mathcal{U}|\!\leq\! |\mathcal{\widetilde{X}}|+2$ for $\mathcal{R}_{1}$ and to $|\mathcal{U}|\!\leq\! |\mathcal{\widetilde{X}}|+1$ for $\mathcal{R}_{2}$. Similarly, let $\mathcal{R}_{3}$ and $\mathcal{R}_{4}$ be the rate regions characterized by (\ref{eq:cscorrkeyrate})-(\ref{eq:cscorrstoragerate}) when $\max\{0,\; I(U;X)-I(U;Y)\}$ is positive and zero, respectively. One can limit the cardinality of $U$ to $|\mathcal{U}|\leq |\mathcal{\widetilde{X}}|+2$ for $\mathcal{R}_{3}$ and to $|\mathcal{U}|\leq |\mathcal{\widetilde{X}}|+1$ for $\mathcal{R}_{4}$.

All regions are convex, and permitting randomization at the encoder and decoder does not improve the regions (see (\ref{eq:fanoapp})$(a)$ and (\ref{eq:storageconv1})$(b)$).

\section{Rate Regions}\label{sec:tightregions}
Suppose we view $\widetilde{X}^n$ as the input to the BC $P_{XY|\widetilde{X}}$. This perspective lets us use known results for the classic BC problem because the channel input $\widetilde{X}^n$ can be viewed as the encoder ``output''. We characterize the key-leakage-storage regions of special classes of BCs $P_{XY|\widetilde{X}}$ for which the rate regions given in Theorem~\ref{theo:correlatedgscs} are tight. 

The key agreement model used in \cite{LaiTrans,IgnaTrans}, which is called the visible source model, corresponds to a semi-deterministic BC such that $P_{XY|\widetilde{X}}\!=\!\mathds{1}\{X\!=\!\widetilde{X}\}P_{Y|\widetilde{X}}$, where $\mathds{1}\{\cdot\}$ is the indicator function. The source model considered in \cite{bizimMMMMTIFS,bizimKittipongTIFS} corresponds to a physically-degraded (PD) BC such that $P_{XY|\widetilde{X}}=P_{X|\widetilde{X}}P_{Y|X}$. 

We show that the regions given in Theorem~\ref{theo:correlatedgscs} are tight, i.e., they are the key-leakage-storage regions, for the following cases:
\begin{enumerate}
	\item a PD BC such that $P_{XY|\widetilde{X}}=P_{Y|\widetilde{X}}P_{X|Y}$, which corresponds to the case where the identifier measurements at the decoder are \emph{better} than at the encoder. There are three different Markov chain conditions that can lead to different rate regions due to the asymmetry in the privacy constraint (\ref{eq:privacyconst}). Note that we consider here a PD BC that has not been studied before. 
	\item LN BCs $P_{XY|\widetilde{X}}$ such that either $I(U;X)\geq I(U;Y)$ or $I(U;Y)\geq I(U;X)$ for all $P_{U|\widetilde{X}}$ where $U-\widetilde{X}-XY$ forms a Markov chain.
\end{enumerate}

\begin{remark}
	\normalfont The rate regions depend on the joint conditional probability distribution $P_{XY|\widetilde{X}}$ rather than only the marginal distributions $P_{X|\widetilde{X}}$ and $P_{Y|\widetilde{X}}$. Thus, the key-leakage-storage regions for the stochastically-degraded BCs are not necessarily equal to the regions for the corresponding PD BCs, unlike in the classic BC problem.
\end{remark}
\begin{remark}
	\normalfont Since $P_{\widetilde{X}XY}$ is fixed, the distinctions between the LN BCs and essentially-less noisy (ELN) BCs is not needed. Observe that the class of ELN BCs is a proper superset of the class of LN BCs \cite[Claim 2]{NairEssentially}.
\end{remark}

The key-leakage-storage regions are as follows.
 
\begin{theorem}[PD BCs]\label{theo:correlatedgscsdegraded}
	Suppose $\widetilde{X}-Y-X$ forms a Markov chain. We have  
	\begin{align}
	&\mathcal{C}_{\text{gs}} = \mathcal{R}_{2}, \qquad\mathcal{C}_{\text{cs}} = \mathcal{R}_{4}.
	\end{align}
\end{theorem}

\begin{theorem}[LN BCs] \label{theo:correlatedessentiallygscsbothcases}
	Suppose $P_{XY|\widetilde{X}}$ is a LN BC with $I(U;X)\geq I(U;Y)$ for all $P_{U|\widetilde{X}}$. We have
	\begin{align}
	&\mathcal{C}_{\text{gs}} = \mathcal{R}_{1},\qquad\mathcal{C}_{\text{cs}} = \mathcal{R}_{3}.\label{eq:regionLNELNCase1gscs}
	\end{align}
	Suppose $P_{XY|\widetilde{X}}$ is a LN BC with $I(U;Y)\geq I(U;X)$ for all $P_{U|\widetilde{X}}$. We have
	\begin{align}
	&\mathcal{C}_{\text{gs}} = \mathcal{R}_{2},\qquad\mathcal{C}_{\text{cs}} = \mathcal{R}_{4}.\label{eq:regionLNELNCase2gscs}
	\end{align}
\end{theorem}

%%%%%%%%%%%%%%%%%%%%%%%%%%%%%%%%
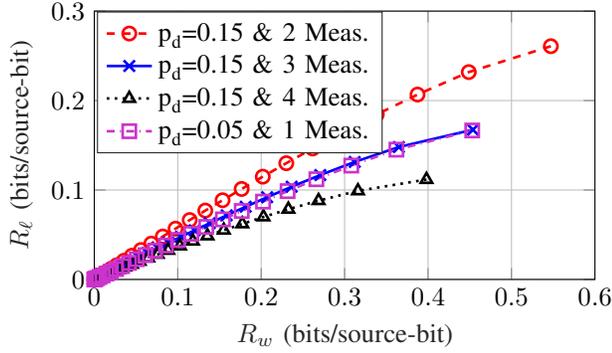
\begin{figure}[t]
	\centering
	\newlength\figureheight
	\newlength\figurewidth
	\setlength\figureheight{9.27cm}
	\setlength\figurewidth{4.8cm}
	\input{New_StorageLeakage.tikz}
	\caption{Storage vs. privacy-leakage rate projection of the boundary triples for the GS model with $\displaystyle p_{\text{e}}\!=\!0.05$.} \label{fig:rw_rl_examplerateregion}
\end{figure} 
%%%%%%%%%%%%%%%%%%%%%%%%%%%%%%%%%%%%%%

The achievability proofs of Theorems~\ref{theo:correlatedgscsdegraded} and \ref{theo:correlatedessentiallygscsbothcases} below follow from Theorem~\ref{theo:correlatedgscs}. In Section~\ref{sec:convessentiallylessnoisycases}, we give the proofs of converses for the BC rate regions in (\ref{eq:regionLNELNCase1gscs}). The proofs of the converses for Theorem~\ref{theo:correlatedgscsdegraded} and the rate regions in (\ref{eq:regionLNELNCase2gscs}) follow similarly and are omitted.

\begin{remark}
	\normalfont The regions $\mathcal{R}_{2}$ and $\mathcal{R}_{4}$ provide strong privacy, i.e., $I(X^n;W)$ vanishes when $n\rightarrow\infty$.
\end{remark}

%%%%%%%%%%%%%%%%%%%%%%%%%%%%%%
	%%%%%%%%%%%%%%%%%%%%%%%%%%%%%%%%
	\begin{figure}[t]
		\centering
		\setlength\figureheight{9.97cm}
		\setlength\figurewidth{4.8cm}
		\input{New_StorageKey.tikz}
		\caption{Storage vs. secret-key rate projection of the boundary triples for the GS model with $\displaystyle p_{\text{e}}\!=\!0.05$.} \label{fig:rw_rs_examplerateregion}
	\end{figure}
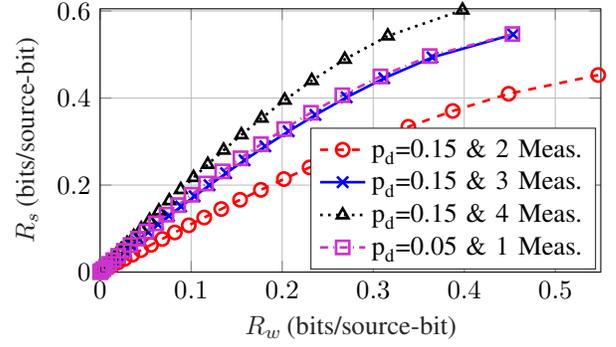 
	%%%%%%%%%%%%%%%%%%%%%%%%%%%%%%%%%%%%%%

%%%%%%%%%%%%%%%%%%%%%%%%%%%%%%%%%%% 

\section{Example}\label{sec:example}
We now compare two different decoder-measurement scenarios to address the practicality of PUF designs for multiple measurements. Consider the GS model with noisy measurements of uniformly-distributed and hidden ring oscillator (RO) PUF outputs $X^n\sim\text{Bern}^n(\frac{1}{2})$ through binary symmetric channels (BSCs) $P_{\widetilde{X}|X}$ and $P_{Y|X}$ after output quantization. 

Since an encoder measurement is made by a manufacturer, the temperature and voltage effects can be eliminated at the encoder, so we assume an encoder measurement through a BSC$(0.05)$, i.e., the crossover probability is $p_\text{e}=0.05$. For the decoder measurement, we assume that cheaper hardware is used to decrease the cost. We model this first case as multiple independent decoder measurements each through a BSC$(0.15)$, where the crossover probability is $p_\text{d}=0.15$. The measurement delay increases linearly with each additional measurement and with the number of ROs used.

As a second case, we can alternatively provide additional protection against environmental effects, which increases the hardware cost approximately linearly with the number of ROs used \cite{bizimtemperature}. We model this second case as a single decoder measurement through a BSC$(0.05)$, i.e., $p_{\text{d}}=0.05$. 

We consider the multiple-measurement case with two, three, and four decoder measurements through a BSC$(0.15)$, and we compare this with a single-measurement through a BSC$(0.05)$. We remark from \cite[Theorem 3]{bizimMMMMTIFS} that it suffices to consider only the conditional probability distributions $P_{\widetilde{X}|U}$ that are BSCs to characterize the rate regions for these channels. We plot the projections of $\mathcal{R}_{1}$ for these scenarios onto the storage vs. privacy-leakage rate plane $(R_{w},R_l)$ in Fig.~\ref{fig:rw_rl_examplerateregion} and onto the storage vs. secret-key rate plane $(R_{w},R_{s})$ in Fig.~\ref{fig:rw_rs_examplerateregion} to compare the effects of linear delay and cost increases. Every marker on each curve corresponds to evaluating the rate-region boundaries for a fixed crossover probability $P_{\widetilde{X}|U}$, so Figs.~\ref{fig:rw_rl_examplerateregion} and \ref{fig:rw_rs_examplerateregion} should be considered jointly. 

Fig.~\ref{fig:rw_rl_examplerateregion} shows that \emph{two} BSC$(0.15)$ decoder measurements yield up to $19\%$ less secret-key rate and up to $80\%$ greater privacy-leakage rate than a single BSC$(0.05)$ decoder measurement. However, \emph{four} BSC$(0.15)$ decoder measurements result in better rate tuples than a single BSC$(0.05)$ decoder measurement, whereas \emph{three} BSC$(0.15)$ measurements achieve similar rate points as a single BSC$(0.05)$. Since additional costs and delays increase linearly with the number of ROs, the fair comparison is between the cases with \emph{two} BSC$(0.15)$ measurements and \emph{one} BSC$(0.05)$ decoder measurement. Thus, if hardware cost is not a bottleneck, one should improve the system design as in, e.g., \cite{bizimtemperature} to obtain a better decoder-measurement channel $P_{Y|X}$ rather than measuring multiple times at the decoder.

%%%%%%%%%%%%%%%%%%%%%%%%%%%%%%%%%%%%%%%%%%%%%%%%%%%%%%%%%%%%%%%%%%%%%%%%%%%%%%%
\section{Proof of Theorem~\ref{theo:correlatedgscs}}\label{sec:achinner}
We provide a proof that follows from the output statistics of random binning (OSRB) \cite{OSRBAmin} method by applying the steps in  \cite[Section 1.6]{BlochLectureNotes2018}. 

%%%%%%%%%%%%%%%%%%%%%%%%%%%%%%%%%%%%%%%%%%%%%%%%%%%%%%%%%%%%%%%%%%%%%%%%%%%%%%%
\subsection{Proof for the GS Model}

\begin{IEEEproof}[Proof Sketch] Fix a $\displaystyle P_{U|\widetilde{X}}$ and let $(U^n,\widetilde{X}^n,X^n,Y^n)$ be i.i.d. according to $P_{U\widetilde{X}XY}=P_{U|\widetilde{X}}P_{\widetilde{X}}P_{XY|\widetilde{X}}$. Assign three random bin indices $(S,W,C)$ to each $u^n$, where $S$ is the secret key, $W$ is the helper data, and $C$ is a public index. Assume $S\in[1:2^{nR_s}]$, $W\in[1:2^{nR_w}]$, and $C\in[1:2^{nR_c}]$.

The reliability constraint (\ref{eq:reliabilityconst}) is satisfied by using a Slepian-Wolf (SW) \cite{SW} decoder to estimate $U^n$ from $(C,W,Y^n)$ if \cite[Lemma 1]{OSRBAmin}
\begin{align}
	R_c+R_w> H(U|Y). \label{eq:slepianwolfdecoder}
\end{align}
	
The strong secrecy and key uniformity constraints in  (\ref{eq:secrecyconst}) and (\ref{eq:uniformityconst}), respectively, are satisfied if \cite[Theorem 1]{OSRBAmin}
\begin{align}
	R_s+R_w+R_c< H(U)\label{eq:independenceofindices}
\end{align}
since (\ref{eq:independenceofindices}) ensures that the three random indices $(S,W,C)$ are almost mutually independent and uniformly distributed.
	
Similarly, the public randomness $C$ is almost independent of $\widetilde{X}^n$, so it is almost independent of $(\widetilde{X}^n,X^n,Y^n)$, if we have \cite[Theorem 1]{OSRBAmin}
\begin{align}
	R_c<H(U|\widetilde{X}).\label{eq:independenceofcode}
\end{align} 
To satisfy the constraints (\ref{eq:slepianwolfdecoder})-(\ref{eq:independenceofcode}), we fix the rates to
\begin{align}
	&R_s = I(U;Y)-2\epsilon\label{eq:chooseR_s}\\
	&R_w = I(U;\widetilde{X})-I(U;Y)+2\epsilon\label{eq:chooseR_w}\\
	&R_c = H(U|\widetilde{X})-\epsilon\label{eq:chooseR_c}
\end{align}
for some $\epsilon>0$ such that $\epsilon\rightarrow0$ when $n\rightarrow\infty$.

{Using the selection lemma \cite[Lemma 2.2]{Blochbook}, there exists a binning that achieves all rate tuples given in (\ref{eq:chooseR_s})-(\ref{eq:chooseR_c}) with strong secrecy when $n\rightarrow\infty$ for the GS model. 
	
	Consider the privacy leakage for this binning.} We have
\begin{align}
	&I(X^n;W|C) \leq H(W)-H(W,C|X^n)+H(C).\label{eq:privacyleakagefirstpartach}
\end{align} 
	
We need to consider two different cases.
	
\textbf{Case 1}: Suppose we have
\begin{align}
	R_c+R_w>H(U|X) \label{eq:case1}
\end{align}
i.e., $H(U|Y)\geq H(U|X)$. Then we can recover $U^n$ from $(C,W,X^n)$ by using a SW decoder \cite[Lemma 1]{OSRBAmin}.

Using (\ref{eq:privacyleakagefirstpartach}), we have for Case 1 that
\begin{align}
&I(X^n;W|C)\nonumber\\
&\overset{(a)}{\leq} n(I(U;\widetilde{X})-I(U;Y)+2\epsilon)-H(W,C|X^n)\nonumber\\
&\qquad+n(H(U|\widetilde{X})-\epsilon)\nonumber\\
& \overset{(b)}{\leq} n(H(U|Y)+\epsilon)-H(W,C,U^n|X^n)+n\epsilon_n\nonumber\\
&\overset{(c)}{\leq} n(H(U|Y)-H(U|X)+\epsilon+\epsilon_n)\nonumber\\
&=n(I(U;X)-I(U;Y)+\epsilon+\epsilon_n)\label{eq:achprivleakcase1}
\end{align} 
where $(a)$ follows by (\ref{eq:chooseR_w}) and (\ref{eq:chooseR_c}), $(b)$ follows from Fano's inequality applied to estimate $U^n$ from $(C,W,X^n)$, given the constraint in (\ref{eq:case1}), for some $\epsilon_n>0$ such that $\epsilon_n\rightarrow 0$ when $n\rightarrow\infty$, and $(c)$ follows because $(U^n,X^n)$ is i.i.d.

\textbf{Case 2}: Suppose we have
\begin{align}
	R_c+R_w<H(U|X)\label{eq:case2}
\end{align}
i.e., $H(U|Y)<H(U|X)$. Then $C$ and $W$ are almost independent given $X^n$ by \cite[Theorem 1]{OSRBAmin}. We remark that \cite[Theorem 1]{OSRBAmin} provides an upper bound on the average variational distance between the joint distribution of $(C,W)$ given $X^n$ and the ideal joint distribution where $C$ and $W$ are independent given $X^n$. The upper bound vanishes when $n\rightarrow\infty$.
			
Using (\ref{eq:privacyleakagefirstpartach}), we have for Case 2 that
\begin{align}
	&I(X^n;W|C)\nonumber\\
	&\overset{(a)}{\leq} H(W)-(H(W|X^n)+H(C|X^n)-\epsilon_n^{\prime})+H(C)\nonumber\\
	&\overset{(b)}{\leq} H(W)-H(W|X^n)-(H(C)-\epsilon_n^{\prime\prime})+H(C)+\epsilon_n^{\prime}\nonumber\\
	&\overset{(c)}{\leq} H(W)-(H(W)-\epsilon_n^{\prime\prime\prime})+\epsilon_n^{\prime}+\epsilon_n^{\prime\prime}\nonumber\\
	&=\epsilon_n^{\prime}+\epsilon_n^{\prime\prime}+\epsilon_n^{\prime\prime\prime}\label{eq:achprivleakcase2}
\end{align} 
where\\
$(a)$ follows {since $W$ and $C$ are almost independent given $X^n$} for some $\epsilon_n^{\prime}>0$ such that $\epsilon_n^{\prime}\rightarrow0$ when $n\rightarrow\infty$,\\
$(b)$ follows because $C$ is almost independent of $X^n$ due to (\ref{eq:independenceofcode}) for some $\epsilon_n^{\prime\prime}>0$ such that $\epsilon_n^{\prime\prime}\rightarrow0$ when $n\rightarrow\infty$,\\
$(c)$ follows because $R_w<H(U|X)$ for Case 2 so that $W$ is almost independent of $X^n$ for some $\epsilon_n^{\prime\prime\prime}>0$ such that $\epsilon_n^{\prime\prime\prime}\rightarrow0$ when $n\rightarrow\infty$. 

{These prove the achievability of the key-leakage-storage region $\mathcal{R}_{2}$ when $n\rightarrow \infty$.}
\end{IEEEproof}

%%%%%%%%%%%%%%%%%%%%%%%%%%%%%%%%%%%%%%%%%%%%%%%%%%%%%%%%%%%%%%%%%%%%%%%%%%%%%%%
\subsection{Proof for the CS Model}
\begin{IEEEproof}[Proof Sketch] We use the achievability proof for the GS model. Suppose the key $S'$, generated as in the GS model together with the helper data $W'$ and public index $C'$, have the same cardinality as the corresponding embedded secret key $S$, i.e., $|\mathcal{S}'|=|\mathcal{S}|$. The chosen key $S$ is independent of $(X^n,\widetilde{X}^n,Y^n)$. Consider the encoder $f_{2}^{(n)}$ with inputs $(\widetilde{X}^n,S)$ and output $W=(S'+S,W')$. Suppose the decoder $g^{(n)}$ with inputs $(Y^n,W)$ and output $\hat{S}=S'+S-\hat{S}'$, where all addition and subtraction operations are modulo-$|\mathcal{S}|$. The decoder of the GS model is used to obtain $\hat{S}'$. 

We have the error probability
\begin{align}
&\Pr[S\ne\hat{S}]=\Pr[S'\ne\hat{S}']\label{eq:errorprobabilityachtheo2}
\end{align}
which is small due to the proof of achievability for the GS model.

Using (\ref{eq:chooseR_s}) and (\ref{eq:chooseR_w}), and from the one-time padding operation applied above, we can achieve a storage rate of
\begin{align}
	&R_w \geq I(U;\widetilde{X})\label{eq:storagecs}
\end{align}
for the CS model.

{Using the selection lemma, there exists a binning that achieves all rate tuples given in (\ref{eq:chooseR_s}), (\ref{eq:chooseR_c}), and (\ref{eq:storagecs}) with strong secrecy when $n\rightarrow\infty$ for the CS model. 

Consider the secrecy leakage for this binning. We obtain}
\begin{align}
	&I(S;W|C')= I(S;W'|C')+I(S;S'+S|W',C')\nonumber\\
	&\overset{(a)}{=} H(S'+S|W',C') - H(S'|W',C')\nonumber\\
	&\overset{(b)}{\leq} nR_s-H(S'|C')+I(S';W'|C')\nonumber\\
	& \overset{(c)}{\leq} nR_s-(nR_s-\epsilon_n^{(4)})+I(S';W'|C')\nonumber\\
	&\overset{(d)}{\leq}\epsilon_n^{(4)}+\epsilon_n^{(5)}
\end{align}
where $(a)$ follows because $S$ is independent of $(W',C',S')$, $(b)$ follows because $|\mathcal{S}'|=|\mathcal{S}|$, $(c)$ follows because $S'$ and $C'$ are almost independent and uniformly distributed due to (\ref{eq:independenceofindices}) for some $\epsilon_n^{(4)}>0$ such that $\epsilon_n^{(4)}\rightarrow0$ when $n\rightarrow\infty$, and $(d)$ follows because the GS model satisfies the strong secrecy constraint (\ref{eq:secrecyconst}) due to (\ref{eq:independenceofindices}) for some $\epsilon_n^{(5)}>0$ such that $\epsilon_n^{(5)}\rightarrow0$ when $n\rightarrow\infty$.

{Consider the privacy leakage for the binning that achieves the rate tuples given in (\ref{eq:chooseR_s}), (\ref{eq:chooseR_c}), and (\ref{eq:storagecs}).} We obtain
\begin{align}
	&I(X^n;W|C')\nonumber\\
	&\leq I(X^n;W'|C') +H(S+S'|W',C')\nonumber\\
	&\qquad -H(S+S'|X^n,W',C',S')\nonumber\\
	&\overset{(a)}{\leq} I(X^n;W'|C')+\log (|\mathcal{S}|)-H(S)\nonumber\\
	&\overset{(b)}{=} I(X^n;W'|C') \label{eq:ach2privleaktemp}
\end{align}
where $(a)$ follows because $S$ is independent of $(X^n,W',S',C')$ and $|\mathcal{S}'|=|\mathcal{S}|$, and $(b)$ follows from the uniformity of $S$. We therefore have the following results for two different cases.

\textbf{Case 1}: Assume that $H(U|Y)\geq H(U|X)$ for the fixed $P_{U|\widetilde{X}}$. By combining (\ref{eq:achprivleakcase1}) and (\ref{eq:ach2privleaktemp}), we obtain
\begin{align}
	I(X^n;W|C')\!\leq\! n(I(U;X)-I(U;Y)+\epsilon+\epsilon_n).
\end{align}

Using the selection lemma, there exists a binning that achieves all rate tuples $(R_s,R_{\ell},R_w)$ in the key-leakage-storage region $\mathcal{R}_{3}$ with strong secrecy when $n\rightarrow\infty$. 

\textbf{Case 2}: Suppose $H(U|Y)<H(U|X)$ for the fixed $P_{U|\widetilde{X}}$. By combining (\ref{eq:achprivleakcase2}) and (\ref{eq:ach2privleaktemp}), we have
\begin{align}
	&I(X^n;W|C')\leq\epsilon_n^{\prime}+\epsilon_n^{\prime\prime}+\epsilon_n^{\prime\prime\prime}. 
\end{align}	

{These prove the achievability of the key-leakage-storage region $\mathcal{R}_{4}$ when $n\rightarrow \infty$.}
\end{IEEEproof}
\subsection{Cardinality Bounds for Theorem~\ref{theo:correlatedgscs}}

Using the support lemma \cite{CsiszarKornerbook2011}, the auxiliary random variable $U$ should have $|\mathcal{\widetilde{X}}|-1$ elements to preserve $\displaystyle P_{\widetilde{X}}$, and three more to preserve $H(X|U)$, $\displaystyle H(\widetilde{X}|U)$ and $\displaystyle H(Y|U)$ for both theorems when $I(U;X)-I(U;Y)>0$. When $I(U;X)-I(U;Y)\leq 0$, as in $\mathcal{R}_{2}$ and $\mathcal{R}_{4}$, we do not need to preserve the term $H(X|U)$.

%%%%%%%%%%%%%%%%%%%%%%%%%%%%%%%%%%%%%%%%%%%%%%%%%%%%%%%%%%%%%%%%%%%%%%%%%%%%%%%
%%%%%%%%%%%%%%%%%%%%%%%%%%%%%%%%%%%%%%%%%%%%%%%%%%%%%%%%%%%%%%%%%%%%%%%%%%%%%%%
\section{Converses for the BC Rate Regions in (\ref{eq:regionLNELNCase1gscs})}\label{sec:convessentiallylessnoisycases}
We use the following lemma to bound the secret-key, privacy-leakage, and storage rates for a class of LN BCs.
\begin{lemma}[\hspace{1sp}\cite{ChandraLessNoisy}]\label{lem:Chandraineq}
	\normalfont For all LN BCs $P_{XY|\widetilde{X}}$ with $I(U;X)\geq I(U;Y)$ for all $P_{U|\widetilde{X}}$ and a fixed $P_{\widetilde{X}}$, we have
	\begin{align}
	&I(S,W,Y^{i-1};Y_{i}) \leq I(S,W,X^{i-1};Y_{i})\label{eq:Chandraineq}
	\end{align} 
	for $i=1,2,\ldots,n$ if $(S,W)-\widetilde{X}^n-(X^n,Y^n)$ forms a Markov chain.
\end{lemma}

Let $U_{i}\triangleq (S,W,X^{i-1})$, which satisfies the Markov chain $U_i-\widetilde{X}_i-(X_i,Y_i)$ for all $i=1,2,\ldots,n$.
%%%%%%%%%%%%%%%%%%%%%%%%%%%%%%%%%%%%%%%%%%%%%%%%%%%%%%%%%%%%%%%%%%%%%%%%%%%%%%% 
\subsection{Proofs of Converses}
Suppose for some $\delta_n\!>\!0$ and $n$, there is a pair of encoders and decoders such that (\ref{eq:reliabilityconst})-(\ref{eq:storageconst}) are satisfied for all LN BCs $P_{XY|\widetilde{X}}$ with $I(U;X)\geq I(U;Y)$ for all $P_{U|\widetilde{X}}$ and a fixed $P_{\widetilde{X}}$ by some key-leakage-storage tuple $(R_{s}, R_\ell,R_{w})$. Using (\ref{eq:reliabilityconst}) and Fano's inequality, we obtain
\begin{align}
H(S|W,Y^n)\!\overset{(a)}{\leq}\!H(S|\hat{S})\!\leq\!n\epsilon_n \label{eq:fanoapp} 
\end{align}
where $(a)$ permits randomized decoding, $\epsilon_n\!=\!\delta_n R_{s} \!+\!H_b(\delta_n)/n$, where $H_b(\delta) = -\delta\log \delta - (1-\delta)\log(1-\delta)$ is the binary entropy function, and $\epsilon_n\!\rightarrow\!0$ if $\delta_n\!\rightarrow\!0$. 

\emph{Secret-key Rate}: We obtain for the GS and CS models
\begin{align}
&n(R_s-\delta_n)\overset{(a)}{\leq} H(S)-H(S|W,Y^n)+n\epsilon_n\nonumber\\
&\overset{(b)}{\leq} I(S;Y^n|W)+n\epsilon_n+\delta_n\nonumber\\
&\leq \sum_{i=1}^n \Big[I(S,W,Y^{i-1};Y_i)+\epsilon_n+\frac{\delta_n}{n}\Big]\nonumber\\
&\overset{(c)}{\leq}\sum_{i=1}^n \Big[I(S,W,X^{i-1};Y_i)+\epsilon_n+\frac{\delta_n}{n}\Big]\nonumber\\
&\overset{(d)}{=}\sum_{i=1}^n \Big[I(U_i;Y_i)+\epsilon_n+\frac{\delta_n}{n}\Big]\label{eq:secretkeyconv1}
\end{align}
$(a)$ follows by (\ref{eq:uniformityconst}) and (\ref{eq:fanoapp}), $(b)$ follows by (\ref{eq:secrecyconst}), $(c)$ follows from Lemma~\ref{lem:Chandraineq}, and $(d)$ follows from the definition of $U_i$.

\emph{Storage Rate}: Observe for the GS model that 
\begin{align}
&n(R_{w}+\delta_n)\overset{(a)}{\geq} H(W|Y^n)+I(W;Y^n)\nonumber\\
&\overset{(b)}{\geq} H(S,W,Y^n)-H(Y^n)-H(S|W,Y^n)\nonumber\\
&\qquad -H(S,W|\widetilde{X}^n)+I(W;Y^n)\nonumber\\
&\overset{(c)}{\geq} I(S,W;\widetilde{X}^n)-I(S,W;Y^n)-n\epsilon_{n}\nonumber\\
&\overset{(d)}{=}\!\sum_{i=1}^{n}[I(S,W,\widetilde{X}^{i-1};\widetilde{X}_i)\!-\!I(S,W,Y^{i-1};Y_i)\!-\!n\epsilon_{n}]\nonumber\\
&\overset{(e)}{\geq}\sum_{i=1}^{n}[I(S,W,X^{i-1};\widetilde{X}_i)\!-\!I(S,W,X^{i-1};Y_i)\!-\!n\epsilon_{n}]\nonumber\\
&\overset{(f)}{=}\sum_{i=1}^{n}[I(U_i;\widetilde{X}_i)\!-\!I(U_i;Y_i)\!-\!n\epsilon_{n}]\label{eq:storageconv1}
\end{align}
where $(a)$ follows by (\ref{eq:storageconst}), $(b)$ follows from the encoding step, $(c)$ follows by (\ref{eq:fanoapp}), $(d)$ follows because the source and channel are memoryless, $(e)$ follows from the data-processing inequality applied to the Markov chain $X^{i-1}-(S,W,\widetilde{X}^{i-1})-\widetilde{X}_i$ and from Lemma~\ref{lem:Chandraineq}, and $(f)$ follows from the definition of $U_i$.

Similarly, observe for the CS model that 
\begin{align}
&n(R_{w}+\delta_n)\overset{(a)}{\geq} I(S,W;\widetilde{X}^n)-H(S|W)+H(S,W|\widetilde{X}^n)\nonumber\\
&\overset{(b)}{\geq} I(S,W;\widetilde{X}^n)+I(S;W)\nonumber\\
&\overset{(c)}{\geq}\!\sum_{i=1}^{n}I(S,W,\widetilde{X}^{i-1};\widetilde{X}_i)\nonumber\\
&\overset{(d)}{\geq}\sum_{i=1}^{n}I(S,W,X^{i-1};\widetilde{X}_i)\nonumber\\
&\overset{(e)}{=}\sum_{i=1}^{n}I(U_i;\widetilde{X}_i)\label{eq:storageconv2}
\end{align}
where $(a)$ follows by (\ref{eq:storageconst}), $(b)$ follows because $\widetilde{X}^n$ is independent of $S$, $(c)$ follows because the source and channel are memoryless, $(d)$ follows by applying the data-processing inequality to the Markov chain $X^{i-1}-(S,W,\widetilde{X}^{i-1})-\widetilde{X}_i$, and $(e)$ follows from the definition of $U_i$.

\emph{Privacy-leakage Rate}: We have for the GS and CS models that
\begin{align}
&n(R_l+\delta_n)\overset{(a)}{\geq} H(W|Y^n)-H(W|X^n)\nonumber\\
&\geq H(S,W,Y^n)-H(S|W,Y^n)-H(Y^n)-H(S,W|X^n)\nonumber\\
&\overset{(b)}{\geq} I(S,W;X^n)-I(S,W;Y^n)-n\epsilon_n\nonumber\\
&\overset{(c)}{\geq}\sum_{i=1}^{n}[I(S,W,X^{i-1};X_i) -I(S,W,X^{i-1};Y_i)	-\epsilon_n]\nonumber\\
&\overset{(d)}{=}\sum_{i=1}^{n}[I(U_i;X_i) -I(U_i;Y_i)	-\epsilon_n]\label{eq:privleakconv1}
\end{align}
where $(a)$ follows by (\ref{eq:privacyconst}), $(b)$ follows by (\ref{eq:fanoapp}), $(c)$ follows 
because the channel and source are memoryless and from Lemma~\ref{lem:Chandraineq}, $(d)$ follows from the definition of $U_i$.

Introduce a uniformly distributed time-sharing random variable $\displaystyle Q\!\sim\! \text{Unif}[1\!:\!n]$ independent of other random variables. Define $X\!=\!X_Q$, $\displaystyle \widetilde{X}\!=\!\widetilde{X}_Q$, $\displaystyle Y\!=\!Y_Q$, and $U\!=\!(U_Q,\!Q)$ so that $\displaystyle U\!-\!\widetilde{X}\!-\!X\!-\!Y$ forms a Markov chain. The converse for all LN BCs $P_{XY|\widetilde{X}}$ with $I(U;X)\geq I(U;Y)$ for all $P_{U|\widetilde{X}}$ and a fixed $P_{\widetilde{X}}$ for the GS model follows by using the introduced random variables in (\ref{eq:secretkeyconv1}), (\ref{eq:storageconv1}), and (\ref{eq:privleakconv1}), and letting $\delta_n\rightarrow0$. Similarly, the converse for the same class of channels for the CS model follows by using the introduced random variables in (\ref{eq:secretkeyconv1}), (\ref{eq:storageconv2}), and (\ref{eq:privleakconv1}), and letting $\delta_n\rightarrow0$.

\section*{Acknowledgment}
O. G\"unl\"u was supported by the German Research Foundation (DFG) through project KR 3517/9-1. O. G\"unl\"u and R. F. Schaefer are now supported by the German Federal Ministry of Education and Research (BMBF) within the national initiative for ``Post Shannon Communication (NewCom)'' under the Grant 16KIS1004. G. Kramer was supported by an Alexander von Humboldt Professorship endowed by the BMBF. O. G\"unl\"u thanks Kittipong Kittichokechai for useful discussions. 

\bibliographystyle{IEEEtran}
\bibliography{IEEEabrv,references}

\end{document}

%% file: New_StorageLeakage.tikz
% This file was created by matlab2tikz.
%
%The latest updates can be retrieved from
%  http://www.mathworks.com/matlabcentral/fileexchange/22022-matlab2tikz-matlab2tikz
%where you can also make suggestions and rate matlab2tikz.
%
\definecolor{mycolor1}{rgb}{0.800000,0.200000,0.800000}%
\begin{tikzpicture}

\begin{axis}[%
width=6.656cm,
height=3.57cm,
at={(0cm,0cm)},
scale only axis,
xmin=0,
xmax=0.6,
xlabel style={font=\color{white!15!black}},
xlabel={$R_{w}$ (bits/source-bit)},
ymin=0,
ymax=0.3,
ylabel style={font=\color{white!15!black}},
ylabel={$R_\ell$ (bits/source-bit)},
axis background/.style={fill=white},
xmajorgrids,
ymajorgrids,
legend style={at={(0.0054,0.4701)}, anchor=south west, legend cell align=left, align=left, draw=white!15!black}
]

\addplot [color=red, dashed, line width=1pt, mark size=2.5pt, mark=o, mark options={solid, red}]
  table[row sep=crcr]{%
0.547165366987767	0.260768409871811\\
0.448801465028878	0.231826682931424\\
0.387401877670431	0.206718575518214\\
0.338574985876171	0.184471361218847\\
0.29740356250059	0.164500142955492\\
0.261662897706234	0.14641967965266\\
0.230102334028755	0.12995982108792\\
0.201931348089566	0.114922274349471\\
0.176611376259109	0.101156376615432\\
0.153755816550012	0.0885445477152275\\
0.133076278888366	0.0769930702291903\\
0.114351267225199	0.0664259869626065\\
0.0974067834670439	0.0567809156197137\\
0.0821037260964734	0.0480060923875414\\
0.0683293823108095	0.0400582313793643\\
0.0559915008800468	0.0329009423862457\\
0.0450140551780347	0.0265035410101981\\
0.0353341499988722	0.020840141239666\\
0.0268997250757069	0.0158889558469073\\
0.019667828278505	0.0116317529095791\\
0.0136033063291062	0.00805343202595066\\
0.00867780902234183	0.00514169421083799\\
0.00486903482253587	0.00288678673821086\\
0.00216016742562131	0.0012813094083195\\
0.000539468134449828	0.000320072552040918\\
0	0\\
};
\addlegendentry{$\text{p}_{\text{d}}\text{=0.15 \& 2 Meas.}$}

\addplot [color=blue, line width=1pt, mark size=3pt, mark=x, mark options={solid, blue}]
  table[row sep=crcr]{%
0.453799921282423	0.167402964166467\\
0.365031897478457	0.148057115381003\\
0.312072870795001	0.131389568642785\\
0.270834713085516	0.116731088428192\\
0.236573759366569	0.103670339821471\\
0.207174464128277	0.0919312460747023\\
0.181459740342924	0.0813172274020892\\
0.158691318384256	0.0716822446441611\\
0.138369301214431	0.0629143015707542\\
0.120136646154763	0.0549253773199788\\
0.103728157210869	0.0476449485516935\\
0.0889409224638513	0.0410156422012584\\
0.0756160845373189	0.0349902166899886\\
0.0636270395092443	0.0295294058003124\\
0.0528714931790178	0.0246003422475726\\
0.0432659409124161	0.0201753824186151\\
0.0347417301368605	0.016231215969024\\
0.0272421911641898	0.0127481824049837\\
0.0207205105060052	0.00970974127720559\\
0.0151381340468532	0.00710205867792729\\
0.0104635578382092	0.00491368353505361\\
0.00667140944954936	0.00313529463804552\\
0.00374175265995769	0.00175950457563268\\
0.00165956857115501	0.000780710553853203\\
0.000414380455995067	0.000194984873586157\\
0	0\\
};
\addlegendentry{$\text{p}_{\text{d}}\text{=0.15 \& 3 Meas.}$}

\addplot [color=black, dotted, line width=1pt, mark size=2.5pt, mark=triangle, mark options={solid, black}]
  table[row sep=crcr]{%
0.398117288632489	0.111720331516533\\
0.315994655616175	0.0990198735187209\\
0.268775485870291	0.0880921837180745\\
0.232571207019703	0.0784675823623791\\
0.202770729318231	0.069867309773133\\
0.177352890378191	0.0621096723246159\\
0.155210879127156	0.0550683661863209\\
0.135660446999815	0.0486513732597202\\
0.118244229546087	0.0427892299024107\\
0.102639300967616	0.0374280321328317\\
0.0886082387444804	0.0325250300853051\\
0.0759710086828324	0.0280457284202394\\
0.0645877756871751	0.0239619078398448\\
0.0543478677660609	0.020250234057129\\
0.045162405878606	0.0168912549471609\\
0.0369592206695027	0.0138686621757017\\
0.0296792521023524	0.0111687379345158\\
0.0232739431531638	0.00877993439395769\\
0.0177033196685562	0.00669255043975653\\
0.0129345565952722	0.00489848122634629\\
0.00894089766034045	0.00339102335718489\\
0.00570083826371404	0.00216472345221019\\
0.00319750939406699	0.00121526130974198\\
0.00141821934777603	0.00053936133047422\\
0.000354123258195949	0.000134727675787039\\
1.11022302462516e-16	1.11022302462516e-16\\
};
\addlegendentry{$\text{p}_{\text{d}}\text{=0.15 \& 4 Meas.}$}

\addplot [color=mycolor1, dashdotted, line width=1pt, mark size=2.5pt, mark=square, mark options={solid, mycolor1}]
table[row sep=crcr]{%
	0.452942548187283	0.166545591071327\\
	0.36208440120615	0.145109619108696\\
	0.307972124076248	0.127288821924032\\
	0.266144871095833	0.112041246438509\\
	0.231651686194514	0.0987482666494164\\
	0.20225229179367	0.0870090737400949\\
	0.176690205027417	0.0765476920865819\\
	0.154174435024942	0.067165361284847\\
	0.134168903688643	0.0587139040449665\\
	0.116291076543007	0.0510798077082226\\
	0.100257409031721	0.0441742003725459\\
	0.0858515480532757	0.0379262677906828\\
	0.0729046455454797	0.0322787776981494\\
	0.0612825813996605	0.0271849476907285\\
	0.0508773502454664	0.0226061993140212\\
	0.0416010726616378	0.0185105141678368\\
	0.0333817238739186	0.014871209706082\\
	0.0261600232773144	0.0116660145181083\\
	0.0198871311599037	0.00887636193110408\\
	0.0145229213896689	0.00648684602074301\\
	0.0100346751593588	0.00448480085620323\\
	0.00639608998008556	0.00285997516858172\\
	0.00358653060963404	0.00160428252530903\\
	0.00159047072672991	0.0007116127094281\\
	0.00039708968905261	0.0001776941066437\\
	0	0\\
};
\addlegendentry{$\text{p}_{\text{d}}\text{=0.05 \& 1 Meas.}$}
\end{axis}
\end{tikzpicture}%

%% file: New_StorageKey.tikz
% This file was created by matlab2tikz.
%
%The latest updates can be retrieved from
%  http://www.mathworks.com/matlabcentral/fileexchange/22022-matlab2tikz-matlab2tikz
%where you can also make suggestions and rate matlab2tikz.
%
\definecolor{mycolor1}{rgb}{0.800000,0.200000,0.800000}%
\begin{tikzpicture}

\begin{axis}[%
width=6.656cm,
height=3.503cm,
at={(0cm,0cm)},
scale only axis,
xmin=0,
xmax=0.55,
xlabel style={font=\color{white!15!black}},
xlabel={$R_{w}\!$ (bits/source-bit)},
ymin=0,
ymax=0.605,
ylabel style={font=\color{white!15!black}},
ylabel={$R_{s}\!$ (bits/source-bit)},
axis background/.style={fill=white},
xmajorgrids,
ymajorgrids,
legend style={at={(0.420,0.00995)}, anchor=south west, legend cell align=left, align=left, draw=white!15!black}
]

\addplot [color=red, dashed, line width=1pt, mark size=2.5pt, mark=o, mark options={solid, red}]
  table[row sep=crcr]{%
0.547165366987767	0.452834633012233\\
0.448801465028878	0.409757992429301\\
0.387401877670431	0.370305933247155\\
0.338574985876171	0.333980094969353\\
0.29740356250059	0.300417247297138\\
0.261662897706234	0.269341508704484\\
0.230102334028755	0.24053680068388\\
0.201931348089566	0.213829840267578\\
0.176611376259109	0.189079069100325\\
0.153755816550012	0.166167137721708\\
0.133076278888366	0.144995626224272\\
0.114351267225199	0.125481229812835\\
0.0974067834670439	0.107552937148434\\
0.0821037260964734	0.0911499014109088\\
0.0683293823108095	0.0762198071290597\\
0.0559915008800468	0.0627175998892605\\
0.0450140551780347	0.0506044870974713\\
0.0353341499988722	0.0398471450280978\\
0.0268997250757069	0.0304170856688009\\
0.019667828278505	0.0222901494951955\\
0.0136033063291062	0.0154460992162252\\
0.00867780902234183	0.00986829594400462\\
0.00486903482253587	0.0055434439554084\\
0.00216016742562131	0.00246139375415289\\
0.000539468134449828	0.000614995870348412\\
0	0\\
};
\addlegendentry{$\text{p}_{\text{d}}\text{=0.15 \& 2 Meas.}$}

\addplot [color=blue, line width=1pt, mark size=3pt, mark=x, mark options={solid, blue}]
  table[row sep=crcr]{%
0.453799921282423	0.546200078717577\\
0.365031897478457	0.493527559979722\\
0.312072870795001	0.445634940122584\\
0.270834713085516	0.401720367760008\\
0.236573759366569	0.361247050431158\\
0.207174464128277	0.323829942282442\\
0.181459740342924	0.289179394369711\\
0.158691318384256	0.257069869972888\\
0.138369301214431	0.227321144145003\\
0.120136646154763	0.199786308116957\\
0.103728157210869	0.174343747901769\\
0.0889409224638513	0.150891574574183\\
0.0756160845373189	0.129343636078159\\
0.0636270395092443	0.109626587998138\\
0.0528714931790178	0.0916776962608514\\
0.0432659409124161	0.0754431598568912\\
0.0347417301368605	0.0608768121386455\\
0.0272421911641898	0.0479391038627802\\
0.0207205105060052	0.0365963002385026\\
0.0151381340468532	0.0268198437268473\\
0.0104635578382092	0.0185858477071222\\
0.00667140944954936	0.0118746955167971\\
0.00374175265995769	0.00667072611798658\\
0.00165956857115501	0.00296199260861918\\
0.000414380455995067	0.000740083548803172\\
0	0\\
};
\addlegendentry{$\text{p}_{\text{d}}\text{=0.15 \& 3 Meas.}$}

\addplot [color=black, dotted, line width=1pt, mark size=2.5pt, mark=triangle, mark options={solid, black}]
  table[row sep=crcr]{%
0.398117288632489	0.601882711367511\\
0.315994655616175	0.542564801842004\\
0.268775485870291	0.488932325047295\\
0.232571207019703	0.439983873825821\\
0.202770729318231	0.395050080479497\\
0.177352890378191	0.353651516032528\\
0.155210879127156	0.31542825558548\\
0.135660446999815	0.280100741357329\\
0.118244229546087	0.247446215813347\\
0.102639300967616	0.217283653304104\\
0.0886082387444804	0.189463666368157\\
0.0759710086828324	0.163861488355202\\
0.0645877756871751	0.140371944928303\\
0.0543478677660609	0.118905759741321\\
0.045162405878606	0.0993867835612631\\
0.0369592206695027	0.0817498800998046\\
0.0296792521023524	0.0659392901731536\\
0.0232739431531638	0.0519073518738062\\
0.0177033196685562	0.0396134910759516\\
0.0129345565952722	0.0290234211784283\\
0.00894089766034045	0.020108507884991\\
0.00570083826371404	0.0128452667026324\\
0.00319750939406699	0.00721496938387728\\
0.00141821934777603	0.00320334183199816\\
0.000354123258195949	0.00080034074660229\\
1.11022302462516e-16	-1.11022302462516e-16\\
};
\addlegendentry{$\text{p}_{\text{d}}\text{=0.15 \& 4 Meas.}$}

\addplot [color=mycolor1, dashdotted, line width=1pt, mark size=2.5pt, mark=square, mark options={solid, mycolor1}]
table[row sep=crcr]{%
	0.452942548187283	0.547057451812717\\
	0.36208440120615	0.496475056252029\\
	0.307972124076248	0.449735686841337\\
	0.266144871095833	0.406410209749691\\
	0.231651686194514	0.366169123603213\\
	0.20225229179367	0.328752114617049\\
	0.176690205027417	0.293948929685219\\
	0.154174435024942	0.261586753332202\\
	0.134168903688643	0.231521541670791\\
	0.116291076543007	0.203631877728713\\
	0.100257409031721	0.177814496080916\\
	0.0858515480532757	0.153980948984759\\
	0.0729046455454797	0.132055075069998\\
	0.0612825813996605	0.111971046107722\\
	0.0508773502454664	0.0936718391944028\\
	0.0416010726616378	0.0771080281076695\\
	0.0333817238739186	0.0622368184015875\\
	0.0261600232773144	0.0490212717496556\\
	0.0198871311599037	0.0374296795846041\\
	0.0145229213896689	0.0274350563840315\\
	0.0100346751593588	0.0190147303859726\\
	0.00639608998008556	0.0121500149862609\\
	0.00358653060963404	0.00682594816831023\\
	0.00159047072672991	0.00303109045304428\\
	0.00039708968905261	0.000757374315745629\\
	0	0\\
};
\addlegendentry{$\text{p}_{\text{d}}\text{=0.05 \& 1 Meas.}$}

\end{axis}
\end{tikzpicture}%